\documentclass{PoS}

\title{Status and Plans for the Array Control and Data Acquisition System of the Cherenkov Telescope Array}

\ShortTitle{Array Control of the Cherenkov Telescope Array}

\author{\speaker{I. Oya}$^{a}$, M. Fuessling$^{a}$, U. Schwanke$^{b}$, P. Wegner$^{a}$, A. Balzer$^{c}$, D. Berge$^{c}$, J. Borkowski$^{d}$, J. Camprecios$^{e}$,  S.~Colonges$^{f}$, J. Colomé$^{e}$, C.~Champion$^{f}$, V.~Conforti$^{g}$, F.~Gianotti$^{g}$, T.~Le~Flour$^{h}$, R.~Lindemann$^{a}$, E.~Lyard$^{i}$, M.~Mayer$^{b}$, D. Melkumyan$^{a}$, M. Punch$^{f}$, C. Tanci$^{j}$, T. Schmidt$^{a}$, J.~Schwarz$^{j}$, G.~Tosti$^{k}$, K. Verma$^{a}$, A. Weinstein$^{l}$, S.~Wiesand$^{a}$, R.~Wischnewski$^{a}$, for the CTA Consortium\footnote{Full consortium author list at http://cta-observatory.org}\\
        E-mail: \email{igor.oya.vallejo@desy.de, matthias.fuessling@desy.de}

{\footnotesize
$^{a}$ DESY, Platanenallee 6, D-15738 Zeuthen, Germany; 
$^{b}$ Institut f\"ur Physik, Humboldt-Universit\"at zu Berlin, Newtonstr. 15, D 12489 Berlin, Germany;
$^{c}$ GRAPPA, Anton Pannekoek Institute for Astronomy, University of Amsterdam,  Science Park 904, 1098 XH Amsterdam, The Netherlands; 
$^{d}$ N. Copernicus Astronomical Center, ul. Rabia\'nska 8 87-100 Torun, Poland;
$^{e}$ Institut de Ciencies de l'Espai (IEEC-CSIC), Campus UAB, Torre C5, 2a planta, 08193 Barcelona, Spain;
$^{f}$ APC, AstroParticule et Cosmologie, Universit\'{e} Paris Diderot, CNRS/IN2P3, CEA/Irfu, Observatoire de Paris, Sorbonne Paris Cit\'{e}, 10, rue Alice Domon et L\'{e}onie Duquet, 75205 Paris Cedex 13, France;
$^{g}$ INAF/IASF - Bologna, via Gobetti 101, 40129 Bologna, Italy;
$^{h}$ Laboratoire d'Annecy-le-Vieux de Physique des Particules, Universit\'{e} de Savoie, CNRS/IN2P3, F-74941 Annecy-le-Vieux, France;
$^{i}$ ISDC Data Centre for Astrophysics, Observatory of Geneva, University of Geneva, Chemin d'Ecogia 19, CH-1290 Versoix, Switzerland;
$^{j}$ INAF - Osservatorio Astronomico di Brera, Via Brera 28, 20121 Milano, Italy;
$^{k}$ Universit\'a di Perugia, Dip. Fisica, Via A. Pascoli, 06123 Perugia, Italy;
$^{l}$ Department of Physics and Astronomy, Iowa State University, Zaffarano Hall, Ames, IA 50011-3160, USA}
	
}	
\usepackage[pdftex]{graphicx}
\usepackage{comment}
\usepackage{txfonts}

\usepackage{xspace}
\usepackage{epstopdf, epsfig}
\usepackage{ulem}

\abstract{The Cherenkov Telescope Array (CTA) is the next-generation atmospheric Cherenkov gamma-ray observatory.  CTA will consist of two installations, one in the northern, and the other in the southern hemisphere, containing tens of telescopes of different sizes. The CTA performance requirements and the inherent complexity associated with the operation, control and monitoring of such a large distributed multi-telescope array leads to new challenges in the field of the gamma-ray astronomy. The ACTL (array control and data acquisition) system will consist of the hardware and software that is necessary to control and monitor the CTA arrays, as well as to time-stamp, read-out, filter and store -at aggregated rates of few GB/s- the scientific data. The ACTL system must be flexible enough to permit the simultaneous automatic operation of multiple sub-arrays of telescopes with a minimum personnel effort on site. One of the challenges of the system is to provide a reliable integration of the control of a large and heterogeneous set of devices. Moreover, the system is required to be ready to adapt the observation schedule, on timescales of a few tens of seconds, to account for changing environmental conditions or to prioritize incoming scientific alerts from time-critical transient phenomena such as gamma ray bursts.  This contribution provides a summary of the main design choices and plans for building the ACTL system. 
}

\FullConference{The 34th International Cosmic Ray Conference,\\
		30 July- 6 August, 2015\\
		The Hague, The Netherlands}

\begin{document}

\section{Introduction}

The Cherenkov Telescope Array (CTA; see \cite{CTA_CONCEPT} for a detailed description) is the next generation ground-based very-high-energy gamma-ray observatory, 
planning to deploy about 100 (20) telescopes  on an area of roughly 4 km$^{2}$ (0.4 km$^{2}$) on a southern (northern) 
site. In the southern array, at least 3 types (2 types in the northern array) of imaging atmospheric Cherenkov telescopes (IACTs),
and potentially more, will be deployed. These telescopes correspond to different sizes (with typical reflector diameters of 23 m, 12 m, and 4 m) whose cameras will comprise 1,000$-$10,000 pixels, acquiring data in the kHz domain. The CTA performance requirements and the inherent complexity in operation, control and monitoring of such a large distributed multi-telescope array leads to new challenges 
in designing and developing the CTA control software and data acquisition (ACTL hereafter) system. In the following,
an overview of the main elements of the preliminary design of
the ACTL system is given.

\section{System Overview}

\begin{figure}
  \centering
  \includegraphics[width=0.65\textwidth]{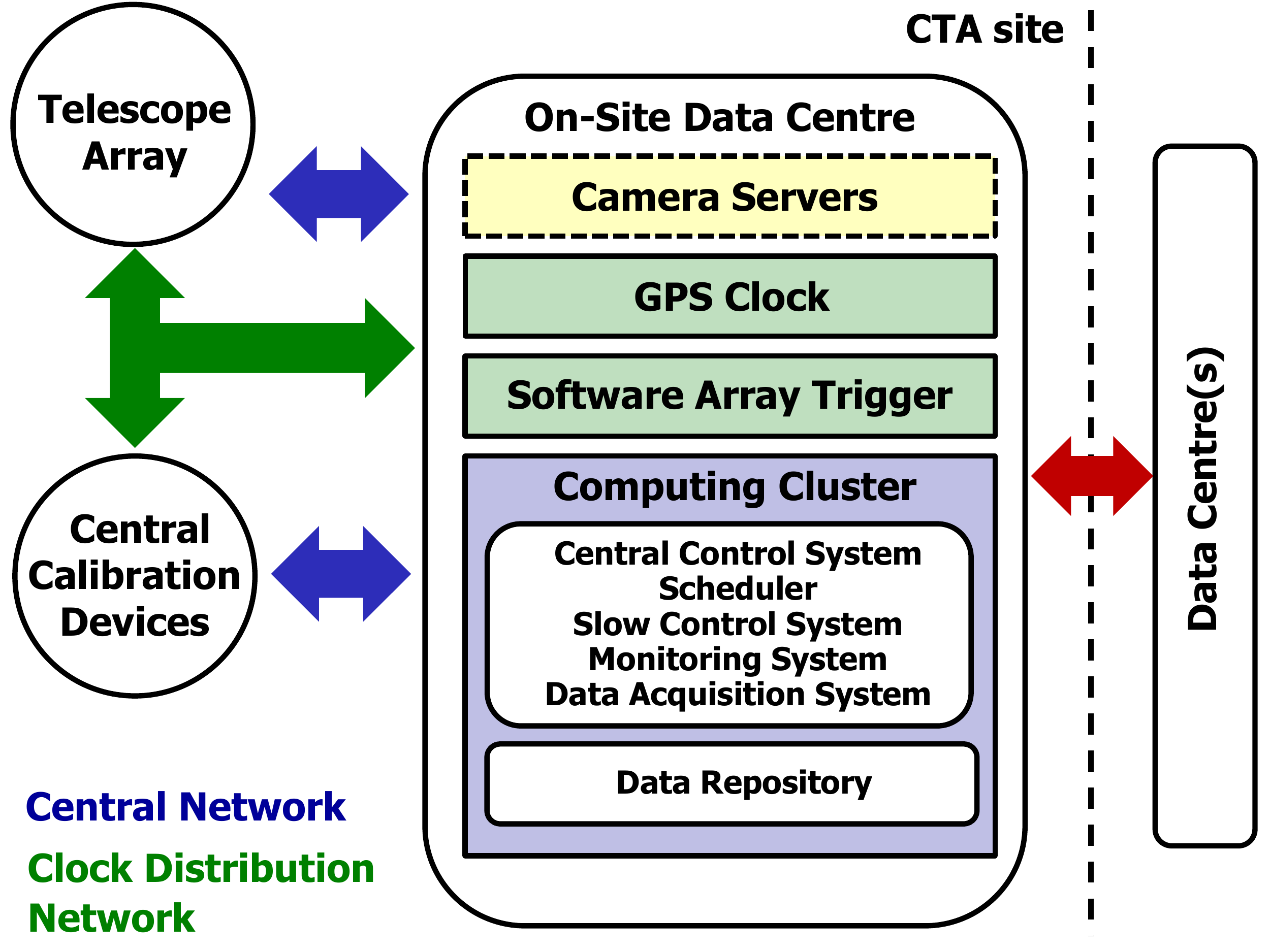}
  \caption{Schematic view of the main components of the ACTL system on a CTA site, as well as the main components that are interfaced (telescopes, central calibration devices, remote data centres).}
  \label{Main_Components}
\end{figure}

CTA operations will be carried out from on-site data centres (OSDC) at the CTA sites. Each OSDC comprises computing
hardware for execution of ACTL software and for mass storage, and is connected to the local CTA array
(for readout and control) and the outside world (for data export). The ACTL work-package will deliver to each CTA installation both the hardware and software components.

The concept for the ACTL hardware emphasises the use of standard, off-the-shelf networking and
computing elements and tries to minimize the amount of hardware that
must be specifically developed for CTA. The main ACTL hardware components will comprise (see Figure \ref{Main_Components}):

\begin{itemize}
\item A single-mode fibre Ethernet wide area network (several 10 Gbit/s) connecting
the telescopes with the OSDC and facilitating data transfer, array-level triggering, and
clock-distribution.
\item Computers (so called camera servers) located in the OSDC and assigned to one telescope to
receive the data after a camera trigger.
\item A central computing cluster (also located in the OSDC) for execution of ACTL software, event
building and filtering, and operation of the data repository. 
\item A White Rabbit (WR) network connecting a central GPS clock with each telescope and the software array trigger (SWAT) and WR interface cards attached to IACT cameras, for time-stamping and array-level triggering, respectively.
\end{itemize}

The design of the ACTL software accounts for the need to
develop, maintain, and operate a more complex and more stable (when compared to existing IACT installations) software system with limited manpower. It prescribes the use of software
frameworks, the application of widely accepted standards, tools and protocols, and follows basically an
open-source approach. The ACTL software is being developed on top of ACS (ALMA Common Software, see \cite{ACS}), a software
framework developed by the European Southern Observatory (ESO) for the implementation of distributed data acquisition and control systems. ACS has been successfully applied in a project of similar scale, ALMA \cite{ALMA}. ACS is based on a
container-component model and supports the programming languages C++, Java and Python, the three languages to be used in CTA, for different purposes. Indeed, C++ will be used for the high performance systems as the data acquisition (DAQ) system (see below), Java for the central and slow control system, and Python for the high level scripts implementing operation modes. The high-level ACTL software will be executed on the central computer cluster and will access most hardware devices via the industry standard {\it Object Linking and Embedding for Process Control Unified Architecture} (OPC UA, see section \ref{slow}). The main ACTL software components will comprise (see Figure~\ref{Main_Components}):

\begin{itemize}
\item A scheduler for CTA (SCTA) optimizing the array use for observations at any point in time given a
list of selected targets, their CTA-assigned scientific priority, the available telescopes, and external
conditions (e.g. weather).
\item A central control system, implementing the execution of
observations using the scheduler and governing the control and coordination the whole array processes, under local supervision by human operators with a graphical user interface (GUI).
\item A slow control system commanding and monitoring each hardware assembly on board of telescopes and central facilities such as weather station.
\item A DAQ system for the telescope Cherenkov data, implementing the further filtering of data
and its storage in the on-site data repository.
\item Monitoring, configuration, logging and alarm services examining and recording instrument and process status data.

\end{itemize}

The ACTL software system will depend on external inputs (e.g. for the long-term scheduling), accept
incoming alerts (e.g. directly from other observatories or in connection with target of opportunities), and manage the
execution (but not the implementation) of a real time analysis software (see \cite{RTA} for details).

\section{The CTA Scheduler}
The SCTA is a flexible scheduler that will be used to automatically create observation
schedules on different time scales (months, nights, hours) in order to ensure that observations
can be performed without human intervention, at high efficiency and optimized for maximum scientific
output of CTA. The off-site service of the scheduler will compute and select the best long-term plan
for a complete season, which will be used afterwards at the observatory sites to compute the short-term
schedule. The off-site service is a time consuming and complex task to be done prior to the proposal
execution on-site. For this purpose, the off-site service will provide a specific GUI to let
operational staff prepare the long term observation plan in an understandable way (which is known to be
a major bottleneck for large infrastructures). Short term plans for the observation during each night will be generated in each OSDC and executed by the central control. The SCTA will be able to react to external and internal triggers and will be tightly coupled with the system to react to changes of the observation conditions and
the state of the array. See \cite{SCHED} for further details.

\section{The Central Control}

\begin{figure}
\centering
\begin{minipage}{0.47\textwidth}
\centering
\includegraphics[width=\textwidth]{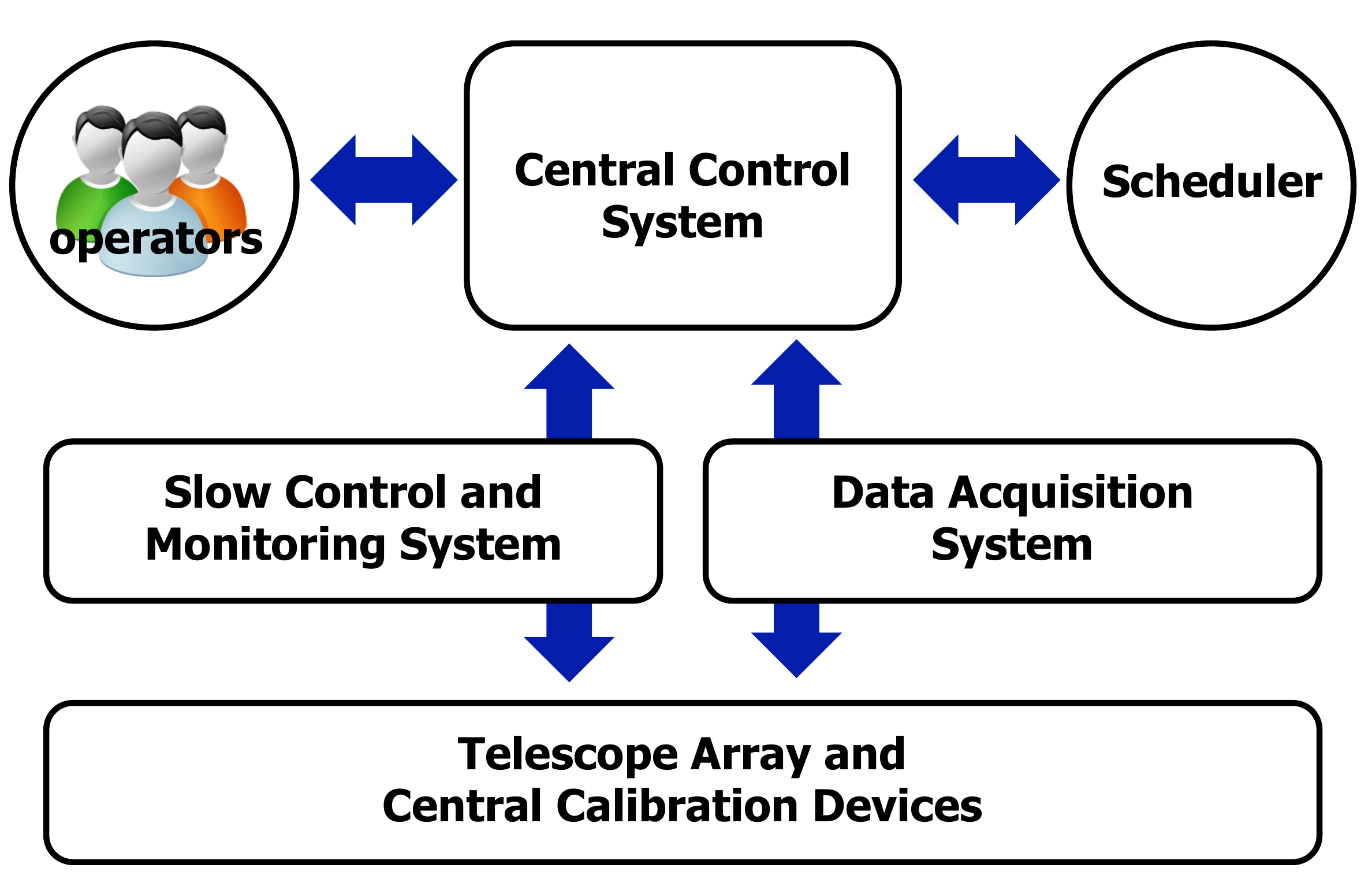}
\end{minipage}\hfill
\begin{minipage}{0.47\textwidth}
\centering
\includegraphics[width=\textwidth]{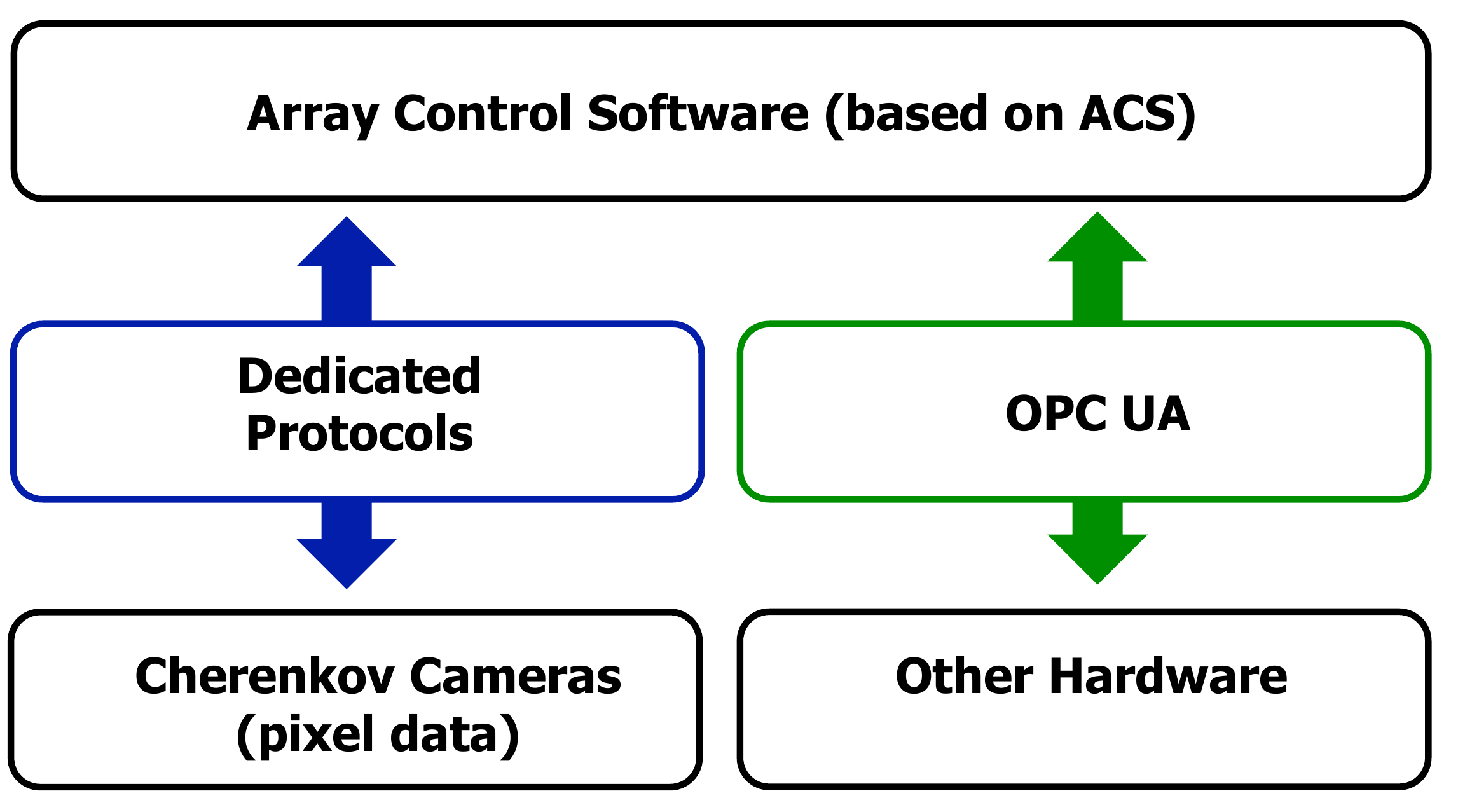}
\end{minipage}
\caption{\label{Figs_2}{\it Right:} Overview of the basic
building blocks of the data acquisition and central array control software. {\it Left:} Schematic wiew of the hardware access in the ACTL system. The ACS-based data acquisition and central
array control software will communicate with most hardware devices via OPC UA (right branch). Due to the high bandwidth
requirements, the readout of the camera data will employ dedicated protocols (left branch).}
\end{figure}

The central control (see the left panel of Figure~\ref{Figs_2}) will interact with all entities in the system, in particular with the SWAT and the individual telescopes to organize the data acquisition for extensive air showers observed by more than one telescope. The central control system monitors the status of the various resources (for example available telescopes at a given moment), and makes this information available to the scheduler for the short term scheduling planning.  It also implements the monitoring of all hardware devices independent of on-going
observations and provides graphical displays and user interfaces. The operator interface design is a particularly challenging element, given that this GUI should allow a crew of just two operators to control and monitor the status of 
an array of 100 telescopes and additional central calibration devices.

The optimized observation schedules provided by the SCTA will be executed by the central control system, which
prepares, starts and stops observations and handles error conditions. The execution of the schedules is performed via high level scripts using the Python language.

\section{Trigger and Time Synchronisation}\label{SWAT}

The cameras in the CTA array will generally trigger locally (i.e. independently
of other telescopes) with an algorithm that suppresses night-sky background light. The array trigger will acts on information sent by the telescope triggers to select shower signatures and reject background processes, allowing a factor of few
data reduction. This array trigger will be implemented via the SWAT software component. It runs on a dedicated
central processor that works in the near real-time regime. The SWAT algorithm operates on a timescale of order 1 second, which gives ample time for any array trigger detection algorithms and compensates for any network induced packet jitters. In the meantime, camera servers buffer the data awaiting the feedback from the SWAT in order to either process the data in the DAQ system, or to discard it.

The ACTL system plans to provide a mechanism to
timestamp data from the various telescopes with nanosecond precision in an array whose extent implies
signal round-trip times exceeding 10 $\mathrm{\mu}$s. The White Rabbit technology \cite{WR_Ref} provides nanosecond accuracy and sub-nanosecond precision
of clock synchronization for large distributed systems. Its aim is to combine the real-time performance
of industrial networks and the flexibility of the Ethernet protocol with the accuracy of dedicated timing
systems. WR is capable of synchronizing more than 1000 nodes, connected by fibres of
up to 10 km in length. 

\section{The Data Acquisition System}

One of the main tasks of the ACTL system will be to transport, process, and store the scientific data acquired with the cameras of the telescopes. 
The data streams generated by the cameras after local camera triggers will dominate the overall data rate of
CTA, with an estimated raw data rate of about 80 GB/s for the southern array. Along the way, inter-telescope and array-level trigger schemes (see section \ref{SWAT}) and possibly an on-line filtering of
events will be used to further suppress the background while selecting electromagnetic showers with
high efficiency.  The DAQ will comprise the camera readout, the buffering
of the read-out data, the processing of array trigger decisions, the building of camera-dependent
events and filtering of interesting events to reduce the overall data volume (about 20 PB/y). Further data volume reduction is planned to be achieved via the data analysis pipeline running on site \cite{RTA}. A compact and flexible data
format is planned to be used since the amount of data to be read out depends strongly on the camera
type which in turn determines parameters like the camera trigger rate, the sampling rate and the decision
as to whether waveform information is being included or not. Further details on the DAQ are presented in \cite{DAQ_ICRC}.

\section{Slow Control}\label{slow}

Besides the Cherenkov camera, each CTA telescope will contain further systems
(drive systems, CCD cameras, sensors, LED flashers, mirror control units etc.) that are needed
for its operation and calibration. The configuration, control, readout, and
monitoring of these devices is taken careof  by the slow control system. In addition there will be devices at array
level (trigger, weather stations, optical telescopes etc.) that must be operated in connection
with the IACTs. In CTA, the device control firmware and telescope onboard computing elements are implemented by means of OPC UA servers. These are planned to be integrated into the ACTL control via {\it bridge} elements implemented as ACS components. In that way, the high-level commands from the central control can be propagated down to the device firmware via the control system, while the monitoring, command execution and alarms are propagated to higher level components via the same framework. See \cite{ACTL_SPIE} for further details on the general approach, and \cite{MST_SPIE, ASTRI_SPIE} for the implementation of the concept in two CTA telescope prototypes.

A set of engineering GUIs, complementary to the operator GUI mentioned above, will allow the display and modification of parameters the normal user/operator should not
need to adjust (e.g. pointing model parameters, servo loop parameters). It is primarily intended for
testing during the commissioning phase of the devices or of a single telescope before it is added to the
CTA array, and to allow for special operation by the observatory maintenance crew.

\section{System Configuration and Monitoring}

A system like CTA will have many hardware and software elements to monitor. These are typically sampled at 1 Hz or slower, or whenever a value changes by more than a predefined threshold. Monitoring
is not restricted to devices but is also applied to the status of
software elements such as the components of the DAQ.
The final number of monitor points for CTA has not yet been determined, but it is expected to be around
10$^{5}$. The effective analysis and interpretation of this large quantity of data depends on maintaining links between
a device's monitor points and the configurations, both current and historical, that determine where
and during what periods the devices was installed and operating in the observatory. For reasons such as these, ACTL will treat the two subjects, configuration and
monitoring, together, via a concept taken from the ALMA Observatory named Telescope Monitoring and Configuration Database (TMCDB). The CTA ACTL team has implemented support for MySQL for the configuration of the TMCDB as a prototype and applied it in
the context of the MST prototype \cite{MST_SPIE}. For storage of the monitoring values, the {\it MongoDB}\footnote{https://www.mongodb.org/} and {\it Cassandra}\footnote{http://cassandra.apache.org/} technologies are currently being evaluated.

Whenever any CTA hardware or software component will detect a situation that is dangerous for the equipment or personnel, an alarm will be raised and then the array operators take the adequate actions. The ACTL team will use the alarm system provided by ACS \cite{ACS_ALARM}, and furthermore, the team is considering the use of complex event processing methods to strengthen the system (see \cite{ESPER} for the use of such systems in IACTs).

\section{On-site Computing and Network Infrastructure}

For each CTA site, the ACTL team is planning to deploy an infrastructure comprising computing/storage hardware, the overall networking
infrastructure (including cabling and switches) and all system services needed (operating system installation, networking services, name services, etc.).

The on-site data centre will host different classes of servers and nodes for computing and storage into the following components, according
to different functionalities divided. The computing nodes are used to deploy the ACTL software components as well as the data analysis procedures described in \cite{RTA}. A total need of about 1500 cores (x86 architecture) for the CTA
    southern array (870 for northern one) is estimated from current software prototypes.The on-site storage system will be entirely assembled from basic storage units consisting
of one storage node server equipped with internally connected disks. Currently the space of a single
(Nearline SAS) disk is 6 TB and the number of disks connected to a head node is 12. Therefore, using
RAID 6 for best redundancy a basic storage system today will provide a net capacity of 60 TB for each node, with a total of 3 PB storage (1.5 PB)
for the southern (northern) installation.

Regarding the array network topology, eight to twenty telescopes will form a group connected to one patch
point or power transformer building in the array. Each telescope is connected via a fibre cable holding 24
single-mode optical fibres, which allows for 12 single connections (two fibres are used for one connection
in RX/TX mode) of up to 10 Gbit/s. Out of these 12 connections, 1 to 6 fibres (depending on the telescope type) will be dedicated to the transfer of the telescope camera data to the camera server located in the data centre. Some additional fibres will be used 
to allow the communication with the drive system, auxiliary devices, management network and slow control. Finally, a fibre will be used for the trigger and timing distribution systems, compatible with the WR solution. By means of a splice sleeve in the patch points the 24-fibre cables of each telescope are connected
to a 216 fibre cable which connects the power transformer building with the data centre.

\section{Conclusions}

An extensive set of solutions has been proposed to organize the ACTL system of the CTA observatory. The proposed solutions are being verified and benchmarked via an intense prototyping activity. Prototypes for the small-sized telescopes (e.g. ASTRI \cite{ASTRI_SPIE} and SST-1M \cite{SST-1M}) and for the medium sized telescope \cite{MST_SPIE} are already being controlled with ACS and OPC UA, while other for ACTL elements such as engineering GUIs and the TMCDB are being tested. Data transfer and serialisation protocols for the DAQ are being prototyped in high-performance computer clusters \cite{DAQ_ICRC}, while equivalent tests are being carried out for the large scale emulation of the monitoring system of the whole array. A set of 15 state of the art computing nodes as well as fiber and a chassis switch, all of them similar to those to be deployed on the OSDCs have been installed at DESY Zeuthen. This tests-bed system is allowing us to emulate the operation of the ACTL hardware and software as a reduced but realistic reproduction of the OSDC environment.

\section*{Acknowledgements}
We gratefully acknowledge support from the agencies and organizations listed under Funding Agencies at this website: http://www.cta-observatory.org/.

\end{document}